\documentclass[prl,twocolumn]{revtex4}
\usepackage{graphicx}
\renewcommand{\v}[1]{{\bf #1}}

\newcommand{\w}{{\omega}}

\def\eqa{\begin{eqnarray}}
\def\eea{\end{eqnarray}}
\newcommand{\eq}{\begin{equation}}
\newcommand{\ee}{\end{equation}}

\newcommand{\<}{\langle}
\renewcommand{\>}{\rangle}
\newcommand{\Tr}{{\rm Tr}}

\renewcommand{\Im}{{\rm Im}}

\newcommand{\ua}{\uparrow}
\newcommand{\da}{\downarrow}
\newcommand{\ra}{\rightarrow}

\newcommand{\Del}{\Delta}

\newcommand{\ga}{\gamma}

\begin{document}

\title{Disentangling single-particle gap by electronic Raman absorption in electron-doped cuprates}
\author{Hong-Yan Lu and Qiang-Hua Wang}
\affiliation{National Laboratory of Solid State Microstructures \&
Department of Physics, Nanjing University, Nanjing 210093, China}


\begin{abstract}
In the under- to optimal-doping regimes of electron-doped
cuprates, it was theoretically suspected that there is a
coexistence of superconducting (SC) and antiferromagnetic (AFM)
orders. The quasi-particle excitations could be gapped by both
orders, and the effective gap is non-monotonic d-wave-like in the
momentum space. Alternatively the gap was also speculated as a
pure pairing gap. Using an effective microscopic model, we
consider the manifestation of the quasi-particle gap in the
electronic Raman spectra in a range of doping levels, where the
relative strength of the SC and AFM order parameters varies. We
demonstrate that from the electronic Raman spectra the effective
single-particle gap can be disentangled into contributions from
the two distinctive orders. This would help to tell whether the
non-monotonic gap is due to the coexistence of SC and AFM orders.
\end{abstract}

 \maketitle

Understanding the pairing symmetry poses a strong constrain on the
underlying superconducting mechanism. In hole-doped cuprates a
consensus is reached that the predominant pairing channel is
$d_{x^2-y^2}$ \cite{hole}. The situation is less clear in
electron-doped cuprates, such as Nd$_{2-x}$Ce$_x$CuO$_4$ and
Pr$_{2-x}$Ce$_x$CuO$_4$. Tunneling \cite{tunneling} and early
Raman measurements \cite{Raman s} suggest an $s$-wave pairing
order parameter, while phase-sensitive measurements consistently
imply $d$-wave pairing\cite{phase sensitive}, the latter of which
is also supported by angle-resolved photo-emission spectra (ARPES)
\cite{ARPES1} and some penetration-depth measurements
\cite{penetration}. It was also speculated that there may be a
transition from $d$- to $s$-wave pairing as electron doping level
increases \cite{ds1} or as temperature decreases\cite{ds2}. Even
in the $d$-wave picture, the situation is more complicated than
that in hole-doped case. The ARPES \cite{ARPES2} and the Raman
scattering \cite{Blumberg} experiments on
Pr$_{0.89}$LaCe$_{0.11}$CuO$_4$ and Nd$_{1.85}$Ce$_{0.15}$CuO$_4$,
both of which are at the optimal doping, revealed a non-monotonic
$d_{x^2-y^2}$-wave quasi-particle gap, {\em i.e.}, the maximal gap
value occurs between the nodal and antinodal directions instead of
at the antinodal direction. On the other hand, a nontrivial
evolution of the Fermi surface (FS) was found in electron-doped
samples \cite{ARPES3,ARPES4}. An electron-like FS pockets exist
near the ($\pm \pi/4$, $\pm \pi$) and ($\pm \pi$, $\pm \pi/4$)
regions in the momentum space for all doping levels in the range
$0.10\leq x \leq0.15$, while a hole-like FS pocket around ($\pm
\pi/2$, $\pm \pi/2$) does not emerge untill $x=0.13$. Accordingly,
Raman scattering experiments revealed that the relative position
of the $B_{1g}$ and $B_{2g}$ peaks changes with
doping.\cite{Blumberg2} In the optimally doped region, the
$B_{2g}$ peak appears at a higher frequency than the $B_{1g}$ one,
while in the overdoped region, it appears at a lower frequency
than the $B_{1g}$ peak. This was argued as due to the change of
the shape of the single-particle gap that originates purely from
superconducting pairing \cite{Blumberg}. This picture is however
not sufficient to account for the evolution of the FS
\cite{ARPES3,ARPES4}, and is also in contradiction to the
tunnelling measurements that consistently report an $s$-wave
feature \cite{tunneling}. Another difficulty, which is perhaps
common in both hole- and electron-doped cuprates, is the large
signal in the $A_{1g}$ channel of the Raman spectra \cite{a1g},
which would have been strongly suppressed by the screening effect
in any one-band theory. A phenomenological two-band theory with
tunable independent pairing gaps on the two bands seems to account
nicely the experimental feature \cite{xiangtao}. Apart from the
$A_{1g}$ puzzle, the microscopic origin of such a theory is yet
lacking. Alternatively, the non-monotonic gap was speculated as
arising from the coexistence of $d$-wave superconducting (SC) and
antiferromagnetic (AFM) orders \cite{yuan,yuan new}, even though
the SC gap itself is in a typical monotonic $d$-wave form. In the
under-doped regime, the hole-pockets do not appear yet, and the
single particle excitations would be gapped everywhere in the
momentum space. This is then not withstanding the $s$-wave feature
in tunnelling\cite{tunneling} and specific heat
measurements\cite{ds2}. An issue does arise, however. ARPES and
tunnelling only probe single particle excitations, it is not clear
how the nontrivial gap manifests in two-particle properties and
whether one can disentangle the single particle gap into
contributions from different origins. This is the topic of the
present paper.

We calculate the electronic Raman spectra in the SC+AFM picture.
In a range of doping levels, the relative strength of AFM and SC
orders varies, and the Raman spectra reveal distinctive peaks that
can be associated to the SC and AFM orders. We also compare the
normal state with AFM alone. It turns out that the $B_{2g}$ Raman
channel is only sensitive to the SC gap, while the $B_{1g}$
channel responses to both SC and AFM orders. In the presence of a
hole-pocket, the $B_{2g}$ channel reveals further a double peak
structure, reminiscent of the two sets of normal state FS.
Moreover, in all cases that we consider, the SC-related Raman peak
in the $B_{2g}$ channel is at lower energy than that in the
$B_{1g}$ channel, although the offset becomes much smaller in the
coexisting SC+AFM state. As an alternative to elastic neutron
scattering measurements (not available yet), these results could
help identify the mechanism of the non-monotonic gap and also
whether SC+AFM coexists in electron-doped cuprates.

As a handy effective model, we consider the two-dimensional
$t$-$t'$-$t''$-$J$ model hamiltonian, \eqa H&=&-t\sum_{\langle ij
\rangle_1\sigma}(c^\dagger_{i\sigma}c_{j\sigma}+{\rm
h.c.})-t'\sum_{\langle ij
\rangle_2\sigma}(c^\dagger_{i\sigma}c_{j\sigma}+{\rm
h.c.})\nonumber\\&&-t''\sum_{\langle ij
\rangle_3\sigma}(c^\dagger_{i\sigma}c_{j\sigma}+{\rm
h.c.})+J\sum_{\langle ij \rangle_1}\vec S_i \cdot \vec S_j,\eea
where $\langle ij \rangle_1$, $\langle ij \rangle_2$, $\langle ij
\rangle_3$ denote the nearest, second-nearest, and third-nearest
neighbor bonds. For electron-doped cuprates, this model is defined
in the hole picture so that $c_\sigma^\dagger$ creates a
$\sigma$-hole, and no double hole occupancy is implicitly assumed.
A hole occupancy of $1-x$ corresponds to a physical electron
doping level $x$. Throughout this work, we use the magnitude of
the nearest neighbor hopping integral as the unit of energy, so
that $t=-1$, $t'=0.32$, $t''=-0.16$, and $J=0.3$. As usual, the no
double occupancy of holes is treated at the slave-boson mean field
level where $c_\sigma$ is replaced by $\sqrt{x}f_\sigma$ in the
hopping terms, where $f_\sigma$ is the fermionic spinon
annihilation operator, and the spin exchange term is decoupled in
a standard fashion as $\vec S_i\cdot \vec S_j\ra
-(3/8)[\chi_{ij}^*\sum_\sigma f_{i\sigma}^\dagger f_{j\sigma}
+\Del_{ij}^*(f_{i\da}f_{j\ua}-f_{i\ua}f_{j\da})+{\rm h.c.}
]+(1/2)(m_j\sum_\sigma \sigma f_{i\sigma}^\dagger f_{i\sigma}+i\ra
j)+(3/8)(|\chi_{ij}|^2+|\Del_{ij}|^2)-m_i m_j$, where consistency
requires $\chi_{ij}=\sum_\sigma\<f_{i\sigma}^\dagger
f_{j\sigma}\>$, $\Del_{ij}=\<f_{i\da}f_{j\ua}-f_{i\ua}f_{j\da}\>$,
and $m_i=\sum_\sigma\sigma \< f_{i\sigma}^\dagger
f_{i\sigma}\>/2$. A different setting was used for the decoupling
of the spin exchange term in ref.\cite{yuan}, although one does
not expect qualitative changes in the results. Anticipating the
uniform AFM order and d-wave SC order, we set $\chi_{ij}=\chi$,
$\Del_{ij}=\pm \Del$ for $x$- and $y$-bonds respectively, and
$m_i=\pm m$ for A and B lattice sites respectively. It is
convenient to redefine $a_\sigma=f_\sigma$ on the A-sublattice and
$b_\sigma=f_\sigma$ on the B-sublattice. The mean field
Hamiltonian of Eq.(1) can then be written as, in the momentum
space, \eqa H_{MF}&=&\sum_{{\v {k}},\sigma}[\varepsilon_{\v
{k}}(a^\dagger_{{\v {k}}\sigma}b_{{\v {k}}\sigma}+{\rm
h.c.})+(\varepsilon_{\v {k}}'-\mu)(a^\dagger_{{\v
{k}}\sigma}a_{{\v {k}}\sigma}+b^\dagger_{{\v {k}}\sigma}b_{{\v
{k}}\sigma}) \nonumber\\&&-2Jm\sigma(a^\dagger_{{\v
{k}}\sigma}a_{{\v {k}}\sigma}-b^\dagger_{{\v {k}}\sigma}b_{{\v
{k}}\sigma})] \nonumber\\&&-\sum_{{\v {k}}}\Delta_{\v {k}}(a_{{\v
{k}}\uparrow}b_{-{\v {k}}\downarrow}+b_{{\v {k}}\uparrow}a_{-{\v
{k}}\downarrow}+{\rm h.c.}),\eea where $\varepsilon_{\v
{k}}=(-2tx-3J \chi/4)(\cos k_x+\cos k_y)$, $\varepsilon_{\v
{k}}'=-4t'x\cos k_x \cos k_y-2t''x(\cos 2k_x+\cos 2k_y)$,
$\Delta_{\v {k}}= -(3J/4)\Delta(\cos k_x-\cos k_y)$. Here $\mu$ is
the chemical potential that fixes the occupancy and the wave
vector $\v {k}$ is restricted in the magnetic Brillouine zone
(MBZ). The order parameters are calculated self-consistently at
each doping levels. For better convenience, we rewrite the mean
field hamiltonian compactly as $H_{MF}=\sum_{\v {k}}\psi_{\v
{k}}^\dagger h_{\v {k}} \psi_{\v {k}}$, where $\psi_{\v
{k}}=(a_{{\v {k}}\uparrow}, b_{{\v {k}}\uparrow}, a_{-{\v
{k}}\downarrow}^\dagger, b_{-{\v {k}}\downarrow}^\dagger)^T$ are
Nambu-Anderson four-spinors, and $h_{\v {k}}$ is a 4 $\times$ 4
single-particle Hamiltonian, \eqa h_{\v {k}}=\left(
\begin{array}{c c c c} {A_1}&{\varepsilon_{\v {k}}}&0&{\Delta_{\v
{k}}}
\\{\varepsilon_{\v {k}}}&{A_2}&{\Delta_{\v {k}}}&0\\0&{\Delta_{\v {k}}}&{A_3}&{-\varepsilon_{\v {k}}}\\
{\Delta_{\v {k}}}&0&{-\varepsilon_{\v {k}}}&{A_4}
\end{array}\right),
\eea where $A_1=\varepsilon_{\v {k}}'-\mu-2Jm$,
$A_2=\varepsilon_{\v {k}}'-\mu+2Jm$, $A_3=-\varepsilon_{\v
{k}}'+\mu-2Jm$, and $A_4=-\varepsilon_{\v {k}}'+\mu+2Jm$. The
single particle hamiltonian $h_k$ can be easily diagonolized,
yielding a set of four eigenvalues $E_{\v k,n}$ ($n=1,\cdots,4$)
and the corresponding eigenvectors $|\v k, n\rangle$ for each $\v
k$. The single-particle $4\times 4$ matrix Matsubara greens
function $G(\v k,i\w_n)=1/(i\w_n{\bf 1}-h_\v k)$, where ${\bf 1}$
is the $4\times 4$ identity matrix, can be expressed in terms of
the eigen states as \eqa G(\v k,i\w_n)=\sum_n\frac{|\v k,n\>\<\v
k,n|}{i\w_n-E_{\v k,n}},\eea which forms the basis for further
calculations.

The electron Raman scattering measures the spectral function of
the fluctuations of the effective charge density $\rho=\sum_{\v
k\sigma}\gamma_\v k c_{\v k\sigma}^\dagger c_{\v k\sigma}$ where
$\gamma_k$ is the Raman vertex that describes the second order
coupling between electrons and photons \cite{literature}. In our
case the bare Matsubara propagator for $\rho$ can be written as,
\eqa \chi_{\gamma\gamma}(i\nu_n)=-\frac {T}{N}\sum_{{\v
k},i\omega_n}\Tr[\gamma_{\v {k}} G({\v k},i\omega_n) \gamma_{\v
{k}}G({\v {k}},i\omega_n+i\nu_n)].\label{generalraman}\eea  Here
$T$ is the temperature, $N$ is the number of lattice sites, and
\eqa \ga_\v k=(\v n_i\cdot\nabla_\v k)(\v n_s\cdot \nabla_\v
k)h_\v k|_{J\ra 0},\eea is the matrix form of the Raman vertex,
where $\v n_i$ and $\v n_s$ are unit vectors for the polarizations
of the incident and scattered lights. Note that the spins do not
couple to light to the leading order, so that the spin exchange
term does not contribute to the Raman vertex, which is reflected
in the formal replacement $J\ra 0$ in the above definition of
$\gamma_\v k$. (We do not consider the higher order two-magnon
Raman absorptions here.) The summation over $i\w_n$ in
Eq.(\ref{generalraman}) can be performed analytically, and we
obtain \eqa \chi_{\gamma\gamma}(i\nu_n)&=&- \frac{1}{N}\sum_{\v
k,m,n} [\frac{f(E_m)-f(E_n)}{i\nu_n+E_m-E_n}\nonumber\\&&
 \times \Tr (\ga_\v k |\v k,m \rangle \langle
\v k,m| \ga_\v k |\v k,n \rangle \langle \v k, n|)],\eea where $f$
is Fermi-Dirac distribution function. The Raman spectral function
is then given by \eqa R(E)=-\frac{1}{\pi}\Im \chi(i\nu_n\ra
E+i0^+),\eea where $E$ is the Raman shift. Note that theoretically
$R(E)$ is odd in $E$ and must vanish at $E=0$. For the $B_{1g}$
and $B_{2g}$ channels, the Raman vertices are given by $\gamma_{\v
k} ^{B_{1g}}=(\gamma_{\v k}^{xx}-\gamma_{\v k}^{yy})/2$ and
$\gamma_{\v k}^{B_{2g}}=\gamma_{\v k}^{xy}$, respectively. The
$B_{2g}$ channel is $d_{xy}$-symmetric and mainly probes
quasi-particle excitations in the nodal directions, while $B_{1g}$
is $d_{x^2-y^2}$ symmetric and mainly probes excitation in the
antinodal directions. These properties enable Raman scattering to
selectively probe excitations in the momentum space, and is
applied in hole-doped cuprates to verify the d-wave nature of the
pairing gap.\cite{literature} Both $B_{1g}$ and $B_{2g}$ vertices
are odd in parity, the Raman absorption in these channels are not
re-normalized by Coulomb interactions. In contrast, in the fully
symmetric $A_{1g}$ channel, the re-normalization is severe and
depends on the detailed band structure (or the Fermi surface
harmonics). Due to such complications, we shall concentrate on the
simpler $B_{1g}$ and $B_{2g}$ channels in this work.

Before the exposition of Raman spectra in the AFM+SC states, it is
instructive to understand, for comparison, the Raman spectra in
the non-superconducting states. First, in the normal metallic
state with neither AFM nor SC, no Raman spectra are anticipated
unless scattering sources are introduced. This is because no
finite-energy particle-hole excitations with zero momentum
transfer can be achieved in ideal metals. Second, in the normal
state with AFM, the band is split into upper and lower bands.
Particle-hole excitations vertically across the sub-bands is
possible, and one expects nonzero Raman absorption. In this case
we show that the Raman response in the $B_{1g}$ channel scales
with the square of the AFM order parameter, while the $B_{2g}$
channel is completely insensitive to AFM for the ranges of doping
under concern. In the AF state, the Hamiltonian in momentum space
can be simplified as $H_{MF}=\sum_{\v {k} \sigma}\eta_{\v
{k}\sigma}^\dagger \xi_{\v k,\sigma} \eta_{\v {k}\sigma}$, where
$\eta_{\v {k}\sigma}=(a_{{\v {k}}\sigma}, b_{{\v {k}}\sigma})^T$,
and $\xi_{\v k,\sigma}=(\varepsilon_{\v
{k}}'-\mu)\sigma_0+\varepsilon_{\v {k}}\sigma_1-2Jm\sigma
\sigma_3$, where $\sigma_0$ is the unit $2\times 2$ matrix and
$\sigma_{1,3}$ are Pauli matrixes. The spin-dependent single
particle Green's function is given by $ g_\sigma(\v
k,i\w_n)=1/(i\omega_n\sigma_0-\xi_{\v {k},\sigma})$, which we
rewrite for later convenience as, \eqa g_\sigma({\v
k},i\omega_n)&=&\frac{1}{2}\sum_{\nu=\pm
1}\frac{1}{i\omega_n+\mu-\varepsilon_{\v {k}}-\nu
E_k}\nonumber\\&&\times(\sigma_0+\nu \frac{\varepsilon_{\v
{k}}\sigma_1-2Jm\sigma \sigma_3}{E_k}), \eea where
$E_k=\sqrt{\varepsilon_{\v {k}}^2+4J^2 m^2}$. The Raman vertex is
given by $\gamma_{\v k}=(n_i\cdot\nabla_\v k)(n_s\cdot\nabla_\v
k)\xi_{\v k,\sigma})|_{J\ra 0}$. By explicit algebra, we obtain
$\gamma_{\v k} ^{B_{2g}}=a_{\v {k}}\sigma_0$, where $a_{\v
{k}}=-4t'x \sin k_x \sin k_y$, and $\gamma_{\v k} ^{B_{1g}}=b_{\v
{k}} \sigma_0+c_{\v {k}}\sigma_1$, where $b_{\v {k}}=4t''x(\cos
2k_x-\cos2k_y)$ and $c_{\v {k}}=t x (\cos k_x-\cos k_y)$. In each
spin-channel, the Raman response is given by a formula similar to
Eq.(\ref{generalraman}). The total response function summed over
spin species is then given by \eqa \chi ^{B_{2g}}&=&\frac{1}{4N}
\sum_{\v {k},\sigma,\nu_1,\nu_2}\frac{f(\nu_2 E_{\v
{k}}+\varepsilon_{\v {k}}-\mu)-f(\nu_1E_{\v {k}}+\varepsilon_{\v
{k}}-\mu)}{i\nu_n+(\nu_1-\nu_2)E_{\v
{k}}}\nonumber\\&&\times\Tr[a_{\v
{k}}(\sigma_0+\nu_1\frac{\varepsilon_{\v {k}}\sigma_1-2Jm\sigma
\sigma_3}{E_{\v {k}}})\nonumber\\&&\times a_{\v
{k}}(\sigma_0+\nu_2\frac{\varepsilon_{\v {k}}\sigma_1-2Jm\sigma
\sigma_3}{E_{\v {k}}}],\eea where $\nu_1, \nu_2=\pm1$. The trace
term gives $2a_{\v k}^2(1+\nu_1\nu_2)$, but only the cases with
$\nu_1=-\nu_2$ contributes eventually due to the two fermi
functions in the first line. Therefore, $\chi ^{B_{2g}}=0$ in the
normal state even with AFM order. On the other hand, for the
$B_{1g}$ channel, the final result is \eqa\chi
^{B_{1g}}&=&\frac{8J^2 m^2}{N} \sum_{\v {k}}[f(-E_{\v
{k}}+\varepsilon_{\v {k}}-\mu)-f(E_{\v {k}}+\varepsilon_{\v
{k}}-\mu)]\nonumber\\&&\times\frac{c_{\v {k}}^2}{E_{\v
{k}}^2}(\frac{1}{i\nu_n+2E_{\v {k}}}-\frac{1}{i\nu_n-2E_{\v
{k}}}),\eea which scales with $m^2$ as we emphasized. We
emphasize, however, that the conclusion that $B_{2g}=0$ depends on
the bare band structure we use. In principle, including 4-th and
further neighbor hopping would cause a nonzero $B_{2g}$, but we
expect it to be much weaker as compared to $B_{1g}$ on general
grounds.

\begin{figure}
\includegraphics[width=7.6cm]{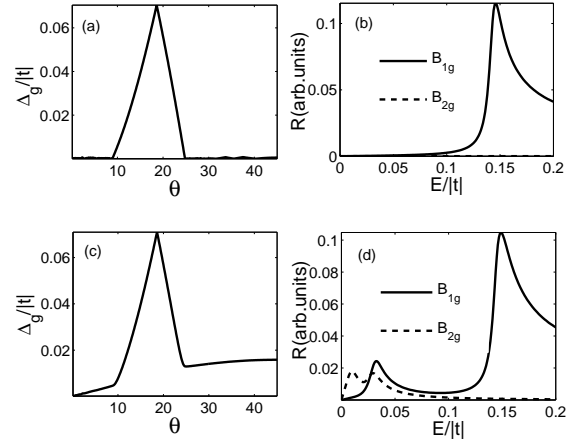}
\caption{Quasiparticle gap $\Delta_g$ and Raman spectra at the
doping level $x=0.143$. (a) $\Delta_g$ as a function of $\theta$
(in degrees) in the AFM normal state. (b) Raman spectra
corresponding to (a). (c) $\Delta_g$ as a function of $\theta$ in
the AFM+SC state. (d) Raman spectra corresponding to (c).}
\end{figure}

We now present and discuss the numerical results. The case of
$x=0.143$ is presented in Figs.1. In order to understand the Raman
spectra we present the quasi-particle gap $\Del_g$ as a function
of the angle $\theta$. Henceforth $\theta$ is defined as the angle
between $\v k-\v Q$ and $-\v Q$ where $\v Q=(\pi,\pi)$. The gap
$\Del_g$ is then defined as the minimum of the quasi-particle
energy in the momentum space along a cut of $\v k$ at the same
angle $\theta$. Clearly $\Del_g(0)$ measures the gap in the nodal
direction, while $\Del_g(45^\circ)$ measures that in the antinodal
region, given the fact that the normal state fermi surface is
close to the antinodal point $(\pi,0)$. We begin with the AFM
normal state (by simply erasing the SC order from a state with
AFM+SC). The gap is presented in Fig.1(a). For $0 \leq \theta \leq
10^\circ$ and $25^\circ < \theta \leq 45^\circ$, the normal state
gap $\Delta_g=0$, roughly reflecting where fermi surfaces appear.
For $10^\circ < \theta \leq 25^\circ$, the normal state is gapped
by the AFM order, with a non-monotonic gap $\Delta_g$ as a
function of $\theta$ due to the underlying band structure. The
Raman spectra in this state are presented in Fig.1(b). As we
analytically proved, for the model at hand, there is no $B_{2g}$
Raman absorption, but the $B_{1g}$ absorption is strong, and
develops a threshold Raman shift roughly twice of the maximal gap
in Fig.1(a).  Now turning on the SC order, the normal state gap
superimposed by the monotonic d-wave pairing gap leads to the gap
structure presented in Fig.1(c). Similar result was also reported
elsewhere.\cite{yuan} The maximum of the gap occurs at
$\theta=18^\circ$. In the experimental case\cite{ARPES2,
Blumberg}, $\theta=20^\circ$ and $\theta=15^\circ$ in $ \rm
Pr_{0.89}LaCe_{0.11}CuO_4$ and $\rm Nd_{1.85}Ce_{0.15}CuO_4$,
respectively. The corresponding Raman spectra are presented in
Fig.1(d). In the $B_{1g}$ channel, the higher energy peak can be
associated to the AFM normal state Raman peak but slightly shifted
due to pairing. The new lower energy peak is definitely caused by
the SC order, as seen from the fact that the energy of this peak
is twice of the pairing gap at the anti-nodal direction. The
$B_{2g}$ channel, which is silent in the normal state, becomes
active in the AFM+SC state, and the peak energy is therefore
entirely determined by pairing alone. Interestingly, because of
the two kinds of fermi surfaces in the case under concern, two
absorption peaks appear in the $B_{2g}$ channel. The energies of
the peaks are roughly (but smaller than) twice of the values of
the pairing gaps at the two fermi surfaces. This is consistent
with the general observation that the $B_{1g}$ Raman vertex is
maximal in the the anti-nodal region, whereas $B_{2g}$ vertex is
maximal in the nodal region and therefore does not see the full
pairing gap. Summarizing, we claim that electronic Raman
scattering not only disentangles the different contributions to
the total single particle gap, but also tells the number of
distinct fermi surfaces, given the monotonic d-wave pairing.
Finally, the SC-driven $B_{2g}$ Raman peaks are at smaller energy
than the SC-driven $B_{1g}$ peak energy, even though the single
particle gap is as non-monotonic as in Fig.1(c). In
experiment\cite{Blumberg}, there appears to be only one peak in
both $B_{1g}$ and $B_{2g}$ channel, and the $B_{2g}$ channel is
situated at a larger energy. It is also not clear whether an
AFM-driven Raman peak is present experimentally.

\begin{figure}
\includegraphics[width=7.6cm]{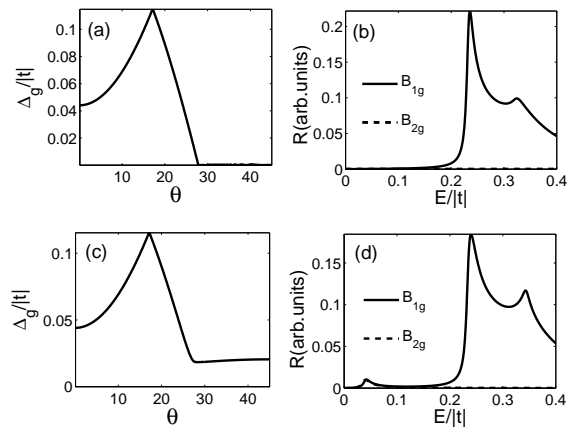}
\caption{The same plot as Fig.1 but for $x=0.13$.}
\end{figure}

We now shift to the under-doped AFM+SC regime where AFM is even
stronger. For $x=0.13$, the AFM normal state single particle gap
is presented in Fig.2(a). Now the gap vanishes only near the
anti-nodal region, reflecting the fact that the hole fermi pocket
is absent and only one kind of fermi surface appear. The
corresponding Raman spectra are presented in Fig.2(b), where again
the $B_{2g}$ channel is silent and the $B_{1g}$ channel develops
AFM-driven peaks. The additional peak at higher energy arise from
the van-Hove singularity in the band structure. Turning on the SC
order, the single particle gap is presented in Fig.2(c). Due to
the underlying AFM normal state gap, the effective gap here is
already nonzero in the nodal direction, and is overall
non-monotonic. The corresponding Raman spectra are presented in
Fig.2(d). In the $B_{1g}$ channel an SC-driven new Raman peak
appears in addition to the AFM-driven peaks at much higher energy.
The new peak is at an energy twice of $\Delta_g=0.02|t|$ at
$\theta = 45^\circ$. On the other hand, the $B_{2g}$ channel shows
a Raman peak with Raman shift very close to but slightly smaller
than the energy of the SC-driven peak in the $B_{1g}$ channel.
Since the $B_{2g}$ channel is silent in the normal state, we
conclude that the gap seen by the $B_{2g}$ channel has nothing to
do with the nodal effective gap, but is rather determined by the
pairing gap. The single-peak structure seen in the $B_{2g}$
channel is also consistent with the the absence of hole pockets. A
disentanglement of single particle gap is therefore again
possible.

\begin{figure}
\includegraphics[width=7.6cm]{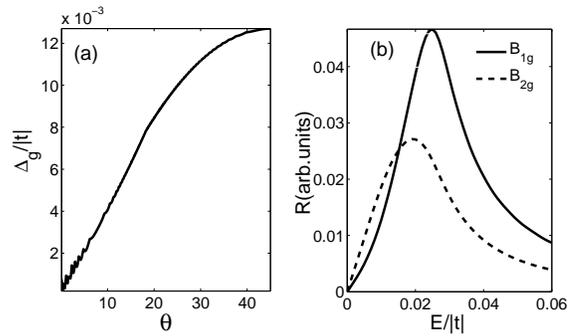}
\caption{(a) $\Delta_g$ as a function of $\theta$ in the SC state
at the doping level $x=0.15$. (b) Raman spectra corresponding to
(a).}
\end{figure}

In the over-doped region, for example at $x=0.15$ in our case, the
self-consistent calculation yields that AFM order disappears, and
we recover a single-band with SC. The situation is then formally
similar to the hole-doped case. For completeness we present the
quasi-particle excitation gap $\Delta_g$ and the corresponding
Raman spectra in Fig.4. The gap as a function of $\theta$ is
typical of a pure d-wave pairing gap. The $B_{1g}$ channel
develops an absorption peak at exactly twice of the maximal
pairing gap, while the ratio is roughly $1.5$ in the $B_{2g}$
channel. Because of the absence of AFM order, both channels
develop single peaks. The relative peak position is in agreement
with experiment\cite{Blumberg2}, but the relative intensity of the
peaks is not, possibly due to more fine details of the materials.
For example, inter-band scattering from the oxygen $p$-band to the
upper Hubbard band under concern may modify the effective mass of
the electrons, leading to photon-energy dependent changes in the
effective Raman vertices. This effect has been experimentally
observed.\cite{Blumberg}

In conclusion, we worked out the consequence of AFM order
coexisting with the SC order in the single-particle gap as well as
the electronic Raman scattering. In particular we show that using
electronic Raman scattering it is possible to disentangle the
effective single particle gap into distinctive contributions from
AFM and SC orders. Combining the studies in the one-band picture
(but with non-monotonic pairing gap) \cite{yuan} and the
phenomenological two-band picture (with independent pairing gaps
on the two bands) \cite{xiangtao}, and by comparing to existing
and forthcoming experiments, a strong constrain can be made on the
AFM origin of the non-monotonic gap as well as the evolution of
the fermi surface with doping in electron-doped cuprates.

\acknowledgments{HYL thanks Jian-Bo Wang for helpful discussions.
This work was supported by NSFC 10325416, the Fok Ying Tung
Education Foundation No.91009, and the Ministry of Science and
Technology of China (973 project No: 2006CB601002).}


\begin{references}
\bibitem{hole} C. C. Tsuei and J. R. Kirtley, Rev. Mod. Phys. {\bf 72}, 969
(2000).
\bibitem{tunneling} Q. Huang, J. F. Zasadzinski, N. Tralshawala, K. E. Gray, D. H. Hinks,
J. L. Peng, and R. L. Greene, Nature {\bf 347}, 369 (1990); Shan
L, Y. Huang, H. Gao, Y. Wang, S. L. Li, P. C. Dai, F. Zhou, J. W.
Xiong, W. X. Ti, and H. H. Wen, Phys. Rev. B {\bf 72}, 144506,
(2005).
\bibitem{Raman s} B. Stadlober, G.Krug, R. Nemetschek, and R. Hackl, Phys. Rev. Lett. {\bf 74},
4911, (1995)
\bibitem{phase sensitive} C. C. Tsuei and J. R. Kirtley, Phys. Rev. Lett. {\bf 85}, 182
(2000); Ariando, D. Darminto, H. -J. H. Smilde, V. Leca, D. H. A.
Blank, H. Rogalla, and H. Hilgenkamp, $ibid$. {\bf 94}, 167001
(2005).
\bibitem{ARPES1} N. P. Armitage, D. H. Lu, D. L. Feng, C. Kim, A. Damascelli, K. M. Shen,
 F. Ronning, and Z.-X. Shen, Phys. Rev. Lett. {\bf 86}, 1126
(2001); T. Sato, T. Kamiyama, T. Takahashi, K. Kurahashi, and K.
Yamada, Science {\bf 291}, 1517 (2001).
\bibitem{penetration} A. Snezhko, R. Prozorov, D. D. Lawrie, R. W. Giannetta, J. Gauthier,
 J. Renaud, and P. Fournier, Phys. Rev. Lett. {\bf 92}, 157005 (2004)
\bibitem{ds1} Amlan Biswas, P. Fournier, M. M. Qazilbash, V. N. Smolyaninova, Hamza Balci,
 and R. L. Greene, Phys. Rev. Lett. {\bf 88}, 207004
(2002); John A. Skinta, Mun-Seog Kim, and Thomas R. Lemberger,
$ibid$. {\bf 88}, 207005 (2002).
\bibitem{ds2} H. Balci and R. L. Greene, Phys. Rev. Lett. {\bf 93}, 067001 (2004).
\bibitem{ARPES2} H. Matsui, K. Terashima, T. Sato, T. Takahashi, M. Fujita, and K. Yamada,
Phys. Rev. Lett. {\bf 95}, 017003, (2005).
\bibitem{Blumberg} G. Blumberg, A. Koitzsch, A. Gozar, B. S. Dennis, C. A. Kendziora, P. Fournier, and R. L. Greene,
Phys. Rev. Lett. {\bf 88}, 107002 (2002).
\bibitem{ARPES3} N. P. Armitage, F. Ronning, D. H. Lu, C. Kim, A. Damascelli,
K. M. Shen, D. L. Feng, H. Eisaki, and Z.-X. Shen, P. K. Mang, N.
Kaneko, and M. Greven, Y. Onose, Y. Taguchi, and Y. Tokura, Phys.
Rev. Lett. {\bf 88}, 257001, (2002).
\bibitem{ARPES4} H. Matsui, K. Terashima, T. Sato, T. Takahashi, S.-C. Wang, H.-B. Yang, H. Ding, T. Uefuji,
and K. Yamada, Phys. Rev. Lett. {\bf 94}, 047005, (2005).
\bibitem{Blumberg2} M. M. Qazilbash, A. Koitzsch, B. S. Dennis, A. Gozar, Hamza Balci, C. A. Kendziora,
R. L. Greene, and G. Blumberg, Phys. Rev. B {\bf 72}, 214510,
(2005).
\bibitem{a1g} For hole-doped cuprates, see, $e.g.$, R.Hackl, W. Gl\"{a}ser, P. M\"{u}ller, D. Einzel, and K. Andres, Phys. Rev. B {\bf 38}, 7133
(1988); T. Staufer, R. Nemetschek, and R. Hackl, P. M\"{u}ller,
H.Veith, Phys. Rev. Lett. {\bf 68}, 1069 (1992); R. Nemetschek, O.
V. Misochko, B. Stadlober, and R. Hackl, Phys. Rev. B {\bf 47},
3450 (1993); for electron-doped ones, see Ref.10 and Ref.13.
\bibitem{xiangtao} C. S. Liu, H. G. Luo, W. C. Wu, and T. Xiang, Phys. Rev. B {\bf 73}, 174517 (2006).
\bibitem{yuan} Qingshan Yuan, Feng Yuan, and C. S. Ting, Phys. Rev. B {\bf 73}, 054501 (2006).
\bibitem{yuan new} Qingshan Yuan, Xin-Zhong Yan, and C. S. Ting, cond-matt/0610523.
\bibitem{literature} See, $e.g.$, T. P. Devereaux, D. Einzel, B. Stadlober, R. Hackl, D. H. Leach, and J. J. Neumeier, Phys. Rev. Lett. {\bf 72}, 396 (1994); T.
P. Devereaux and D. Einzel, Phys. Rev. B {\bf 51}, 16336 (1995);
T. Strohm and M. Cardona,  $ibid$. {\bf 55}, 12725 (1997); T.
Strohm and M. Cardona, Solid State Commun. {\bf 104}, 233 (1997).
\end{references}
\end{document}